# Seismic Activity of the Earth, the Cosmological Vectorial Potential And Method of a Short-term Earthquakes Forecasting

Baurov Yu. A., Baurov A.Yu., Baurov A.Yu. (jr.),
Spitalnaya A.A., Abramyan A.A., Solodovnikov V.A.
Closed joint stock Company Research Institute of Cosmic Physics, 141070,
Moscow region, Korolev, Pionerskaya, 4.

**Abstract**

To the foundation of a principally new short-term forecasting method there has been laid down a theory of surrounding us world's creation and of physical vacuum as a result of interaction of byuons – discrete objects. The definition of the byuon contains the cosmological vector-potential $\mathbf{A}_g$ - a novel fundamental vector constant.

This theory predicts a new anisotropic interaction of nature objects with the physical vacuum.
A peculiar "tap" to gain new energy (giving rise to an earthquake) are elementary particles because their masses are proportional to the modulus of some summary potential $\mathbf{A}_\Sigma$ that contains potentials of all known fields. The value of $\mathbf{A}_\Sigma$ cannot be larger than the modulus of $\mathbf{A}_g$. In accordance with the experimental results a new force associated with $\mathbf{A}_\Sigma$ ejects substance from the area of the weakened $\mathbf{A}_\Sigma$ along a conical formation with the opening of $100^\circ \pm 10^\circ$ and the axis directed along the vector $\mathbf{A}_\Sigma$. This vector has the following coordinates in the second equatorial coordinate system: right ascension $\alpha \approx 293° \pm 10°$, declination $\delta \approx 36° \pm 10°$.

The 100% probability of an earthquake (earthquakes of 6 points strong and more by the Richter scale) arises when in the process of the earth rotation the zenith vector of a seismically dangerous region and/or the vectorial potential of Earth's magnetic fields are in a certain way oriented relative to the vector $\mathbf{A}_g$.

In the work, basic models and standard mechanisms of earthquakes are briefly considered, results of processing of information on the earthquakes in the context of global spatial anisotropy caused by the existence of the vector $\mathbf{A}_g$, are presented, and an analysis of them is given.

## 1. Introduction

In Ref.[1-3] theory of surrounding us world's creation and of physical space (vacuum) as a result of interaction of byuons – discrete objects, is given. The definition of the byuon contains the cosmological vector-potential $\mathbf{A}_g$ - a novel fundamental vector constant.

This theory predicts a new anisotropic interaction of nature objects with physical vacuum. Peculiar "tap" to gain new energy (giving rise to an earthquake) are elementary particles because their masses are proportional to the modulus of some summary potential $\mathbf{A}_\Sigma$ that contains potentials of all known fields. The value of $\mathbf{A}_\Sigma$ cannot be larger than the modulus of $\mathbf{A}_g$. In accordance with the experimental results shown in [1-9], this force ejects substance from the area of the weakened $\mathbf{A}_\Sigma$ along a conical formation with the opening of $100^\circ \pm 10^\circ$ and the axis directed along the vector $\mathbf{A}_\Sigma$. This vector has the following coordinates in the second equatorial coordinate system: right ascension $\alpha \approx 293° \pm 10°$, declination $\delta \approx 36° \pm 10°$ [3,4].

In the present paper, the hypothesis for connection of the Earth's seismic activity with fluctuations in the structure of physical space (vacuum) is considered. That is, our planet, the Earth, will be treated as a peculiar large-scale probe in the physical space the seismic activity of which allows to judge the fluctuations of physical space itself and in so doing to refine its structure and direction of the vector $\mathbf{A}_g$. And vice versa, with a knowledge of space structure and of its fluctuations one would be able to predict more accurately the most dangerous day time for Earth's inhabitants at one or another place of our planet. The urgency of this problem has no need of much to talk because earthquakes lead to many-thousand victims and catastrophic destructions.

In the work, basic models and standard mechanisms of earthquakes are briefly considered, results of processing of information on the earthquakes in the context of global spatial anisotropy caused by the existence of the vector $\mathbf{A}_g$, are presented, and an analysis of them is given. A method of short-term earthquakes forecasting (1-2 days in advance) is advanced.

### 2. Earthquakes and their mechanisms of generation

An earthquake is the sudden release of potential energy of bowels of the earth in the form of shock waves and elastic vibrations propagating every which way from the center (hypocenter), as well as in the form of shifts of the Earth's surface such as warpings of crust, displacements along fault lines, compactions of granular or not grouted precipitations, soil slips and mudflows, soil dilution, snow avalanches, new fracture formation in rocks [10].

The mechanism of release of potential energy is not yet sufficiently studied for earthquakes at depths from 60km to 720km [11].

The Earth is never quiescent. Sensible seismographs constantly detect weak oscillations, microseisms, with periods from 4 to 6 seconds and variable amplitudes. Even "microseismic storms" take place, they are connected mainly with sharp changes in weather. During powerful earthquakes almost total seismic energy is released.

An energy characteristic of earthquakes is the magnitude. A scale of magnitudes was elaborated by Karl Richter, a professor of Californian Technological Institute. The magnitude of an earthquake is a measure of the total energy quantity released in the form of elastic waves, and that is not equal to the total energy of earthquake. Part of elastic energy accumulated in the center of earthquake is turned into heat.

The magnitude is characterized by the maximum amplitude of record by a seismograph of standard type at a fixed distance from the hypocenter of earthquake. The most potent disastrous earthquakes have amplitude $M = 9$. As an example of earthquakes with $M = 9$, that in Lissabon at November 1, 1755, may be considered.

Any earthquake with $M \geq 7$ is the great disaster, particularly if that happens in the neighbourhood of populated area. The energy released during such earthquakes ranges from $2.1 \cdot 10^{22}$ to $1.6 \cdot 10^{25}$ erg.

Very strong earthquakes lead to torsion and spheroidal oscillations of the Earth as a whole. For very weak shocks, negative values of the magnitude are used [10].

As to depth of occurrence (h) of earthquake centers, those are classified into three categories: **crustal** ($0 \leq h \leq 70$km), middle focal (intermediate) ($70 < h \leq 300$km), and deep focal $h > 300$km earthquakes.

The most deep focal shocks were recorded only in the southern hemisphere of the Earth:

- in the vicinity of Kermadec runner (10047 m in depth, interval of latitudes $35°S \leq \varphi \leq 25°S$):

- in the neighbourhood of Tonga runner (10882 m in depth, interval of latitudes $25°S \leq \varphi \leq 15°S$) (both runners are located to the north-west from New Zealand);

- in the latitude $\varphi \approx 6°$ to the south from Sulawesi island (Indonesia).

The centers of those earthquakes were up to 720km deep from the surface.

In Table 1 are listed distributions of earthquakes of all types (from deep focal to crustal ones) registered at a period from 1904 till 1988[1] and having magnitudes $M = 7$ and more except for the deep focal earthquakes for which, to magnify the sample, we had to use those beginning with $M \geq 5.9$. The data were chosen from catalogues [10-15].

It follows from Table 1 that great majority of earthquakes of all type occurs at latitudes 35°-45° north as well as in southern latitudes of equatorial region and at 25°S. Therewith the crustal earthquakes are prevalent in the northern hemisphere, whereas deep focal earthquakes are in the southern one as evidenced by confidence intervals L obtained with the use of Student t-test for random values of the average with the level of significance of 0.01 and 0.001 (see Fig.1).

The first fact was known as early as the beginning of 20-th century [16-19] and even earlier [20] in the context of the question of how and where the stresses accumulate in the Earth's crust.

The question of a particular role of 35°-th parallel in the tectonic life of the Earth was repeatedly posed by geographers and geologists as in the Russian so in foreign literature. The first work along this line was probably that by A.A.Tillo [20] which has pointed out all the most high peaks and mountain chains as well as the most great depths in the northern hemisphere to gravitate towards the latitudes of 30°-40°. All researches of this sort **persued** prior 1962 (with voluminous list of literature) may be acquainted with in the articles of M.V.Stovas, an astronomer-geodesist [16,17], and V.A.Tsaregradsky, an astrogeologist [18]. One of important conclusions of those studies was that all processes connected with the most potent earthquakes have a common mechanism for the whole Globe.

Following German geologists, the authors of Refs. [16-18], considered the periodic variation of the Earth's rotation velocity due to tide influence of Moon and Sun as one of the sources of energy for tectonic processes. The smaller is the angular velocity of rotation, the more globe-shaped becomes the form of the Earth. And vice versa, with the growth of angular velocity the planet becomes more ellipsoid-shaped. The reconstruction of superficial form of the Earth will be the most considerably reflected in the regions of 35°-th parallels of northern and southern hemispheres. Those regions are named critical. Said parallels are also associated with compressive and tensile strains originated under the action of precession (slow motion of the Earth's axis of rotation over a circular cone around an axis orthogonal to the plane of ecliptic. Many epicenters of earthquakes concentrate near the critical parallels, too.

---

[1] Precisely this period of time will be analyzed further without reference to dates.

To explain large number of meridionally extending structures, supporters of rotational hypothesis admitted the possibility of time variations of the Earth's axis of rotation.

As the form of the Earth changes, all tectonic processes are to be symmetric about the equator. But from Figs.1 is clearly seen that at southern latitudes of 30°-40°S the number of the most strong earthquakes is lesser than the average, and there is no symmetry in their distribution about the equator.

In the summary curve of latitude distribution of earthquakes, the maximums are the **sames** as in the curves of their depth-distribution, i.e. $\varphi = 25°S$ and $\varphi = 5°$ on the south, $\varphi = (45°-35°)$ N on the north.

A substantial predominance of crustal earthquakes in the Northern hemisphere ($6.1\sigma$) and deep focal earthquakes in the Southern one ($4.5\sigma$) as well as their statistics equality ($1\sigma$) for the middle focal earthquakes, are also important facts.

In this regard it is interesting to note that the irregularity of distribution of oceans and continents over the Earth's surface has first drawn attention of Ch. Layel who has established an idea of evolution and principles of actualism in geology. In the first edition of his book "Basic principles of geology" published in the years 1830-1833, he gives a projection of Earth's hemispheres which is displaced in such a manner that one of them is mainly continental whereas the other is oceanic. Therewith the "northern pole" is represented by a district of London ($\varphi \sim 51°N$) and the "southern pole" is located close to New Zealand. The plane of "equator" forms at that an angle of 39° with that of geographic equator.

When investigating the planetary jointing of the Earth, it was elucidated [21,22] that the modern global relief presented mainly by lineaments (i.e. by extended straight lines manifested as contours of continental coasts, mountain chains, river valleys, and divides) is considerably a summary effect of manifestation of **crustal** jointing associated with epicenters of earthquakes. Hence the lineaments are very much to do with the most new tectonic displacements of the Earth's crust. The statistic distribution of lineamental azimuths reveals the existence of two quadrupole (cruciform) systems with dominant orientation along the "meridian-parallel" directions as well as diagonally to those [21,22]. The Sun's "zones of instability" (regions of predominant origin of spot groups and flare activity) manifest similar systems of preferable directions. The filaments arranged in parallel with such a zone, are more stable as compared with those of orthogonal orientation, i.e. the filaments themselves also form quadrupole systems of two types. The similarity of orientation of Sun's "instable zones" associated with the most potent non-stationary processes, and of lineaments connected with the present tectonic activity on the Earth, has raised a question of the possibility of existence of some energy factors external to the Solar system and manifesting themselves in the form of the solar activity and tectonic actions on planets and their satellites [23,24]. The results of a number of works [23-30] dedicated to analyse of cosmic factors in the geotectonics and above all to elucidation of the relation between the solar activity (SA) and the seismicity of the Earth, have led to the conclusion that such relation does exist, that the earthquakes occur the most often in a phases of increase and decrease of solar cycle, and that the frequency of earthquake events depends more likely on SA-fluctuations than on its level itself.

Several mechanisms for this relation were proposed:

a) According to Refs. [25-27], it can be accomplished through general atmospheric processes on planet. The intensification of corpuscular radiation of the Sun when an active region passes through the central meridian, increases the potential energy of the atmosphere, depresses zonal circulation, and leads to local accumulation of energy. Redistribution of atmospheric masses over the Globe disturbs the balance of the Earth's shape and results in displacements of crust blocks. The earthquakes are consequences of these displacements.

b) Magneto-dynamic influences of the Sun on the Earth change its speed of rotation [27,31,32] which leads to the above mentioned rotational effect.

c) In the Earth's crust, currents with daily variations are universally present [33]. Additional currents due to SA are meant to sufficiently act on mountain rocks and lead to detonation of earthquakes [27]. Thus the solar flares are considered as possible trigger mechanism of earthquakes in the Earth's areas being in instable state.

d) In Ref. [28] a question is raised whether or not the manifestations of Sun's and Earth's activity are simply events coincident in time, and the cause of them to be beyond as the Earth so the Sun. This question is posed on the basis of a number of common morphologic symptoms appearing in the course of development of chromospheric flares and earthquakes despite the fact that the process goes in plasma in the first case and in the solid body in the second case.

Let us analyze the seismic activity of the Earth in the context of action of the new force.

Consider a mechanism of increase of this force due to Earth's currents. As was said above, there exists in the vicinity of the Earth some real summary potential $\mathbf{A}_\Sigma$ equal to the sum of $\mathbf{A}_g$ and potential from magnetic fields of Galaxy, Sun, Earth, etc.

The new force is of complex nonlinear and nonlocal nature and can be represented as some series in $\Delta A_\Sigma$ [1-3,9]. As a first approximation we have

$$F \sim \Delta A_\Sigma \frac{\partial \Delta A_\Sigma}{\partial x} \qquad (1)$$

where $x$ is the spatial coordinate.

The quantity $|\mathbf{A}_\Sigma|$ is always lesser than $|\mathbf{A}_g|$ (see [1-3]). There are possible fluctuations of enormous magnitudes of vectorial potentials of magnetic sources remote from the Earth, for example, that of the Sun. Despite the large values of $\Delta \vec{A}_\Sigma$ of those sources, magnitudes of $\frac{\partial \Delta \vec{A}_\Sigma}{\partial x}$ from them are, as was said above, as a rule insignificantly small since variations of potentials take place through vast distances. Hence the values of the new force from remote sources are very small, too, and the values of $\frac{\partial \Delta \vec{A}_\Sigma}{\partial x}$ due to currents in the Earth might be several orders greater than those from remote sources. Then the Earth might work as some amplifier of enormous fluctuations of $\Delta \vec{A}_\Sigma$ from remote sources because the magnitude of the new force includes the product $\Delta \vec{A}_\Sigma \frac{\partial \Delta \vec{A}_\Sigma}{\partial x}$ where the first factor is created by remote cosmic sources and the second one is by the currents in the Earth.

Give an example.

In [1-3] the results of an uninterrupted experiment (from *Feb.24* to *Mar.22, 96*) with a tide gravimeter developed in the Sternberg Astronomical Institute of Moscow University on base of a standard quartz gravimeter "Sodin" (Canadian production) were showed (Fig.2,3). To measure the new interaction by the Sodin gravimeter, a constant magnet (*60mm* in diameter, *15mm* in height, the field *B* in the center of *0.3T*) was attached to it in such a way that the vector-potential lines of the magnet in the vicinity of a test platinum weight were directed perpendicular to the Earth's surface (i.e. towards the vertical component of the vector $\vec{A}_G$).

The major deflections of the gravimeter, being repeated every *24 hours*, are associated with the Moon's gravity. Denote an average amplitude of Moon tide by *L*, an amplitude of accidental events, recorded by gravimeter and corresponding to an increase in Moon attraction, by $K L^+$, and that corresponding to a decrease by $K L^-$, where $K$ is a factor indicating the value of deflection in terms of Moon tide amplitudes.

Three events were documented: on the *28th of February*, at $10^{05}$; *4th of March*, at $10^{58}$; *18th of March*, at $20^{54}$. Two last of them had a huge amplitude (*13.6L$^-$* and *15.2L$^-$*, respectively) and $\Delta t \approx 10$ *min*. A time profile of the event on *18th of March 1996* is shown in Fig.3 at a more large time scale.

Here $\Delta A$ was created by some natural sources, and $\frac{\partial \Delta A}{\partial x_1}$ was created by a constant magnet attached to the gravimeter, since $\Delta A_S$ from spatial sources varies at immensely long distances, with an infinitesimal value of $\frac{\partial \Delta A}{\partial x_1}$ at the point of the gravimeter location.

The fact of local change in the gravitation field described in [1-3,34] (see Fig. 2,3) may act as trigger for the stresses accumulated in the Earth's crust, and cause earthquakes. At that the role of amplifier may be played, instead of a constant magnet, by the currents existing inside the Earth.

Further, the maximum mass change in a local characteristic volume of the Earth takes place when this volume passes, during the Earth's rotation, through the region of maximum change in $A_\Sigma$ by the vectorial potential $A_E$ of the Earth. To estimate that, it will suffice to assume the existence of a circular current in the Earth (Fig.4) the magnetic field of which is observed by us as geomagnetism. Near the current the vectorial potential of its magnetic field is directed along the vector of current (Fig.4). Thus if the vector $\mathbf{A}_\Sigma$ in some volume of the Earth experiences the maximum change by $\mathbf{A_E}$, the masses of elementary particles in that same volume are also changed in accordance with the theory of byuons, and hence the total mass of the volume is changed, too. The latter leads to maximum change in the gravitation field of the Earth that causes earthquakes.

Denote the zenith distance (the arc of the great circle of celestial sphere between the direction to zenith and that to the point of intersection of $\mathbf{A}_g$ with the sphere) by $z_{\tilde{A}_g}$. In Table 2 is given the dependence of distribution of $z_{\tilde{A}_g}$ arcs on the latitude $\varphi$ of earthquake epicenters: n is the number of earthquakes, $\bar{n}$ is their average, L – confidence interval, $C_p$ – average number of earthquakes in column at the fixes $z_{\tilde{A}_g}$ calculated for two assumed values of the $\alpha$-coordinate of $\mathbf{A}_g$ being equal to 270° and 290°; $\Sigma_{270°}$, $\Sigma_{290°}$ are the correspondent sums for columns of table. The coordinates $\varphi°$, $z_{\tilde{A}_g}$ for a cells of Table denote:

$n_i$ ($\alpha= 270°$)   $n_i / \bar{n}$ ($\alpha = 270°$)

$n_i$ ($\alpha= 290°$)   $n_i / \bar{n}$ ($\alpha = 290°$)

where $n_i$ is the number of earthquake epicenters in a given interval of latitudes at the moment when apexes of $\mathbf{A}_g$ were in the interval of zenith distances $z_{\tilde{A}_g av} \pm 10°$. The average values $\bar{n} = n/9$ are indicated for each interval of latitudes in 12-th column of Table; $n_i / \bar{n}$ is the relative number of earthquakes normalized to the average $\bar{n}$. For example, at $\varphi = 55°$ and $z_{\tilde{A}_g} =90°$ we have 57 → 4.01 and 46 → 3.24. That is in the interval of latitudes (50°-59°) there are happened 57 ($\alpha= 270°$) and 46 ($\alpha= 290°$) earthquakes when the apex of $\mathbf{A}_g$ was in the interval (80°-99°) of zenith distances $z_{\tilde{A}_g}$ i.e. in the region of sunset. At that $\bar{n} = 128/9 = 14,2$; $4.01 = 57/14.2$ and $3.24 = 46/14.2$.

It is seen from Table 2 that this theoretical statement is in total consistency with experiment. I.e. the maximum of earthquakes is observed at an angle of 90° between the direction to zenith and the vector $\mathbf{A}_g$.

As was above said, the local increase in the Earth's currents gives effects similar to those shown in Figs.2,3 (i.e. currents in the Earth are playing part of magnet attached to the gravimeter), since in that case a considerable enhancement of $\frac{\partial \Delta \dot{A}_\Sigma}{\partial x}$ may occur and, as the result, of the magnitude of new force functioning as trigger for earthquakes.

As was shown in experiments with the plasma generator in gimbal suspension [3,4], the energy of the plasma jet increases up to 40% when the vectorial potential $\mathbf{A}$ of current of the plasma generator has an angle of ~ 135° relative to the vector $\mathbf{A}_g$.

On the basis of experiments with plasma generator and those on investigating the action of the new force on the β-decay rate of radioactive elements [1-3,5], in Fig.4 a cone of directions of action of the new force on the Earth's substance is shown (in the case that there are currents there flowing antiparallel to $\mathbf{A}_g$). As is seen from this Figure, in the Northern hemisphere the maximum of new force will fall into latitudes between 30° – 40° due to the current system of the Earth's magnetic field and the known direction of vector $\mathbf{A}_g$.

The substance at those latitudes should "bulge" from the Earth.

As was said above, in those latitudes we have maximum number of earthquakes of substantionally superficial character. The mountain rocks themselves are of lesser hardness there than in the Southern hemisphere where the new force compresses the substance and rejects the epicenters of earthquakes deeper into the Earth. At that the epicenters become closer to the plane of equator as is shown in Fig.1. The maximum number of earthquakes occurs in southern latitudes, at 25°S and more close to the equator.

The long-term investigations of the influence of the new force on the rate of β-decay of radioactive elements [1-3,5] as well as experiments with plasma devices on energy generation with the aid of that force [1-4] allow to conclude that in the bowels of the earth similar manifestations of new force might take place which lead to local earth heating up, origination of various stresses in its deep layers, and hence to accumulation of potential energy liberated during earthquakes. The interaction of current systems with the physical space through the new force is possibly the main contribution of energy to maintain energy balance of celestial bodies: stars and planets. Such statement is naturally only a hypothesis requiring thorough and long verification.

Thus the following factors point to the fact of influence of new interaction and the global anisotropy caused by the vector $\mathbf{A}_g$, on earth's tectonic processes and particularly on earthquakes:

1. Predominance of crustal earthquakes in Northern hemisphere and of deep focal ones in the Southern hemisphere.
2. Maximum probability of initiation of earthquakes when a local Earth's volume passes the direction drawn from the center of the Earth perpendicularly to the vector $\mathbf{A}_g$ (the region of maximum change in the gravitational field in the result of decreased $|\mathbf{A}_\Sigma|$ under the action of vectorial potential of the Earth's magnetic field). A reverse estimation of direction of $\mathbf{A}_g$ gives therewith its coordinates: $\alpha \approx 270°\text{-}290°$, $\delta \approx 30°$ (from the distribution of earthquakes).
3. Existence of an "oceanic" and a "continental" hemispheres of the Earth.
4. Absence of symmetry in the number of the most potent earthquakes at 35° north and 35° south. The southern hemisphere is formed by more strong mountain rocks than the Northern one.
5. Stresses in the earth's crust experience daily variation along parallels of latitude and meridians (dependence of $z_{\tilde{A}_g}$ on latitude $\varphi$ and on hour angle t). That creates not only a system of parallel and meridional stresses but also a diagonal quadrupole system. The action of those stresses during long periods of time leads to structural rearrangements in the earth's crust. The influence of new interaction on the Earth has to have a near-yearly variation, too, which probably manifests itself in the Chandler's motion of poles and, precisely, in on-earth experiments as well [1,2].
6. The higher is the latitude $\pm\varphi$, the lesser is the probability of potent earthquakes. In the Antarctic there are mountains and fractures but no very strong earthquake was on record in the entire period of observation on that territory.

**3. Method of a Short-term Earthquakes Forecasting**

The analysis of collected data about strong earthquakes (see above) shows that practically all of them happened at the moments when a seismically dangerous region was in a certain manner oriented in space relative to the stars.

To clarify the regularity observed, consider the diagram in Fig.5 that conventionally represents the positions of the Earth and the earthquake zone at the moment of the event (see Table 3). For simplicity, the diagram is brought to the plane of the ecliptic. The Earth 1 (globe) moves relative to the Sun 2 in an orbit 3 lying in the ecliptic plane. The latter is inclined to the equatorial plane at an angle of $\varepsilon = 23°26^{'}$ (in the diagram the plane of the ecliptic and the rotation of the Earth around its axis are shown as circles). Eight positions of the Earth are given at the moments of earthquakes (Positions: 1-8; (Table 3)).

The Earth rotates around its axis, and the regions of earthquakes also move along circular trajectories 4 relative to this axis. At the moments of earthquakes their regions were in positions 1-8. The vectors **V** (Positions: 1-8) of linear velocities of daily rotation of earthquake's regions as well as the vector $\mathbf{A}_g$ take certain positions relative to each other. (Vectors **V** are antiparallel with the vector potential **A** from the magnetic field of the Earth). Namely, if one brings the vector $\mathbf{A}_g$ and vector **V** of linear velocities, by a parallel transfer, to the point of the earthquake epicenter, then in the moment of earthquake the vectors **V** of circular velocities of the earthquake regions fall into the limits of a zone restricted by two conic surfaces circumscribed about the vector $\mathbf{A}_g$, and are directed along the generatrices of the family of conic surfaces inside the zone in consideration. The openings of the cones forming that zone are, respectively, 90° and 110°.

The analysis of additional factors accompanying the earthquakes has shown that the seismic activity increased when the earthquakes were preceded by severe changes in the total activity of the Sun (i.e. changes in $\Delta A_\Sigma$).

From the above reasoning one can indicate, with an accuracy of ± 1 hour, the time points when a seismically dangerous region will go through the most hazardous spatial directions. Will an earthquake happen or not at a given timepoint, show special devices measuring changes in $\partial A_\Sigma/\partial t$.

The detected regularities are used as the basis for a method of short-term forecast of earthquakes with the probability close to 100%.

The invention [35] has received a recognition in the form of three gold medals in International Exhibitions in Brussels, Geneva, and Seoul.

### 4. Conclusion

As distinct from all earlier models of earthquakes [16-33], the theory of byuon allows to establish the deep-seated nature of physics of earthquakes, and therefore to predict place and time of an earthquake with the probability close to 100%.

**References**
1. Baurov Yu.A., Structure of Physical Space and New Method of Obtaining Energy (Theory, Experiment, Applications), Moscow, "Krechet", 1998 (in Russian).
2. Baurov Yu.A., On the structure of physical vacuum and a new interaction in Nature (Theory, Experiment and Applications), Nova Science, NY, 2000.
3. Baurov Yu.A, Global Anisotropy of Physical Space. Experimental and Theoretical Basis. Nova Science, NY, 2004.
4. Yu.A.Baurov, I.B.Timofeev, V.A.Chernikov, S.F.Chalkin, A.A.Konradov., Experimental Investigation of the Distribution of Pulsed-plasma-generator at its Various Spatial Orientation and Global Anisotropy of Space. Phys. Lett. A, V.311, (2003), p.512.
5. Baurov Yu.A., Konradov A.A., Kuznetsov E.A., Kushniruk V.F., Ryabov Y.B., Senkevich A.P., Sobolev Yu.G., Zadorozsny S. Mod. Phys.Lett A. v.16, N 32 (2001), p.2089.
6. Yu.A.Baurov, E.Yu.Klimenko, and S.I.Novikov, Doklady Akademii Nauk (DAN), 1990, v.315, p.1116.
7. Yu.A.Baurov, E.Yu.Klimenko, S.I.Novikov, Phys. Lett. A 1992, v.162 p.32.
8. Yu.A.Baurov, P.M.Ryabov, DAN, 1992, v.326, p.73.
9. Yu.A.Baurov, Phys.Lett. A 1993, v.181, p.283.
10. Aiby J.A. Earthquakes, M, "Progress", 1978 (in Russian).
11. Thomaas C. Nichols, jr. (University of Colorado). Global summary of human response to natural hazards: earthquakes, p.274-284, in coll. work: Natural Hazards Local, National, Global, edited by Gilbert F.White, New York, London, Toronto, 1974.
12. Golubeva N.V. Catalogue of strong earthquakes of the Globe, 1953 to 1967 with M≥6, Acad. Of Sci. USSR, Inst. Of the Earth's physics, M., 1972, 164p. (in Russian).
13. Regional Catalague of Earthquakes, Berkshire, United Kingdom (1972-1981).

14. Catalogue of Epicentres in the International Seismological Summary.
15. Seismic Bulletin of the Network of Reference Seismic Stations of USSR (in Russian).
16. Stovas M.V. On Stressed State of Crustal Layer in Region between 30° and 40°, in coll. work "Problems of Planetary Geology", Gosgeoltechizdat, M., 1963, pp.275-284 (in Russian).
17. Stovas M.V. Several Questions of Tektogenesis, in cool. work "Problems of Planetary Geology", Gosgeoltechizdat, M., 1963, pp.222-274 (in Russian).
18. Tsaregradsky V.A. To Question on Deformations of the Earth's Crust, in cool. work above, Gosgeoltechizdat, M., 1963, pp.149-221 (in Russian).
19. Veronnet *Al.* La forme de la Terre et sa constitution interne, Paris, 1914.
20. Tillo A.A. Average Height of Land and Average Depth of Sea in Northern and Southern Hemispheres. Dependence of Average Depths of Seas on Geographic Latitude. Izvestiya Russkogo Geograficheskogo obshchestva, v.XXV, N 6, 1889 (in Russian).
21. Schulz S.S. (Elder), in coll. work "Basic Problems of Study of Quaternary Period", M., 1965, pp.147-150. (in Russian).
22. Voronov P.S. Essays on Morphometry Laws of Global Relief of Earth, L., Nauka, 1965 (in Russian).
23. Shpitalnaya A.A. On Importance of Choice of Reference System During Investigation of Non-Stationary Cosmic processes, in cool. work "Problems of Investigation of Universe", N 7, M.-L., 1978 (in Russian).
24. Shpitalnaya A.A. To the Question of Destruction of Filaments by Flares. Solar Data N 2, 1971 (in Russian).
25. Sytinsky A.D. Geomagnetisdm and Aeronomy. v. 3, N 1, 1963; v. 6, N 4, 1966 (in Russian).
26. Sytinsky A.D. Uspekhy Phyzicheskikh Nauk (UFN), v.III, N 2, 1973, p.367 (in Russian).
27. Simpson J.F. Science Letters, **3**, 1968, 418.
28. Shpitalnaya A.A., Vasilieva G.Ya., Pystina N.S. On Possibility of Action of Gravitation Waves on Activity of Earth and Sun, in cool. work "Problems of Investigation of Universe", N 4, M.=L., 1975, pp.129-137 (in Russian).
29. Yakovlev B.A. Astronomical collective work of Lvov State University, N 3,4, 1960, p.152 (in Russian).
30. Afanassieva V.N. Geomagnetism and Aeronomy, v. 3, 1963, p.561 (in Russian).
31. Danjon A., CR Acad. Scien., Paris, **254**, 1962, p.3058.
32. Gribbin J., Plagemaun S. Nature, **243**, 1973, p.26.
33. Hessler V.P., Wescott E.M, Nature, **184**, Suppl. 9, 1959, p.627
34. Yu.A.Baurov, A.V.Kopaev, "Experimental Investigation of Signals of a New Nature with the Aid of Two High Precision Stationary Quartz Gravimeters" Hadronic Journal (2002), v.25, p.697.
35. Yu.A Baurov, A.Yu. Baurov, A.Yu. Baurov, A.A. Shpitalnaya, A.A. Abramyan, V.A. Solodovnikov. Method of Short-term forecasting Heightened Seismic Danger of Seismically Dangerous Regions", Invention ¹ 2006110310 RF, March 31, 2006.

Table 1. The dependence of number of earthquakes ($n_i$) from the latitude (or declination) of their epicenters and hypocenters.

$\Sigma$ - is the total number of earthquakes with M ≥ 7 over the period from 1904 till 1988,

$\tilde{n}$ - is the average number of earthquakes in 10°-interval of latitudes,

$\sigma$ - is the standard deviation characterizing the anisotropy of Northern and Southern hemispheres relative to the number of the most strong earthquakes,

$n_s$ - is the number of earthquakes in the southern hemisphere,

$n_N$ - is that in the Northern hemisphere.

| $\varphi°=$ $=\delta°$ | h > 300 êì | | 70< h ≤ 300 êì | | 0 ≤ h ≤ 70 êì | | Âñå | |
|---|---|---|---|---|---|---|---|---|
| | $n_i$ | $n_i/\bar{n}$ | $n_i$ | $n_i/\bar{n}$ | $n_i$ | $n_i/\bar{n}$ | $n_i$ | $n_i/\bar{n}$ |
| +85 | 0 | 0,00 | 0 | 0,00 | 0 | 0,00 | 0 | 0,00 |
| +75 | 0 | 0,00 | 0 | 0,00 | 1 | 0,02 | 1 | 0,01 |
| +65 | 0 | 0,00 | 2 | 0,12 | 14 | 0,25 | 16 | 0,15 |
| +55 | 26 | 0,78 | 17 | 1,05 | 92 | 1,64 | 135 | 1,28 |
| +45 | 69 | 2,06 | 20 | 1,23 | 120 | 2,14 | 209 | 1,98 |
| +35 | 70 | 2,09 | 30 | 1,85 | 122 | 2,18 | 222 | 2,10 |
| +25 | 45 | 1,34 | 16 | 0,99 | 75 | 1,34 | 136 | 1,29 |
| +15 | 13 | 0,39 | 33 | 2,04 | 92 | 1,64 | 138 | 1,31 |
| +5 | 23 | 0,69 | 19 | 1,17 | 85 | 1,52 | 127 | 1,20 |
| -5 | -99 | 2,96 | 42 | 2,59 | 137 | 2,45 | 278 | 2,63 |
| -15 | 70 | 2,09 | 49 | 3,02 | 99 | 1,77 | 218 | 2,06 |
| -25 | 168 | 5,01 | 48 | 2,96 | 68 | 1,22 | 284 | 2,69 |
| -35 | 20 | 0,60 | 9 | 0,56 | 42 | 0,75 | 71 | 0,67 |
| -45 | 0 | 0,00 | 1 | 0,06 | 25 | 0,45 | 26 | 0,25 |
| -55 | 0 | 0,00 | 5 | 0,31 | 20 | 0,36 | 25 | 0,24 |
| -65 | 0 | 0,00 | 0 | 0,00 | 15 | 0,27 | 15 | 0,14 |
| -75 | 0 | 0,00 | 0 | 0,00 | 0 | 0,00 | 0 | 0,00 |
| -85 | 0 | 0,00 | 0 | 0,00 | 0 | 0,00 | 0 | 0,00 |
| $\Sigma$ $\bar{n}$ $n_S$ $n_N$ | 603 33,5 357 246 | 4,5$\sigma$ | 291 16,2 154 137 | $\sigma$ | 1007 55,9 406 601 | -6,1$\sigma$ | 1901 105,6 917 984 | 1,5$\sigma$ |

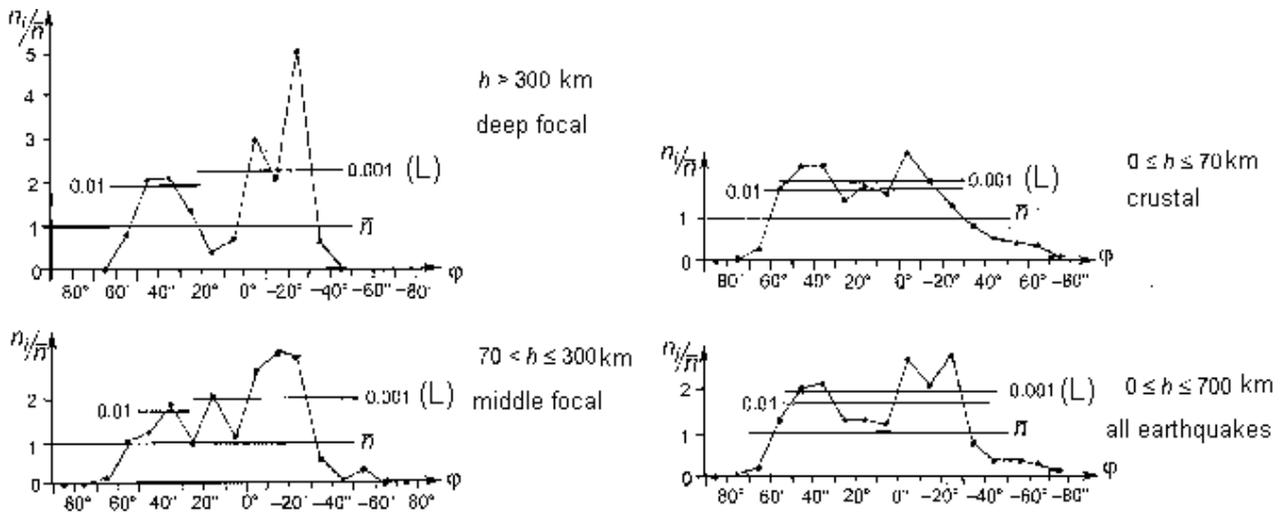

Fig.1
The distribution of earthquakes (in relative units $n_i/\bar{n}$) in dependence from latitude (ö) and depth (h) of earthquakes hypocenter.
$\bar{n}$ – average numbers of earthquakes ($\bar{n}$ =1); L – confidence interval at significance level 0,01-0,001

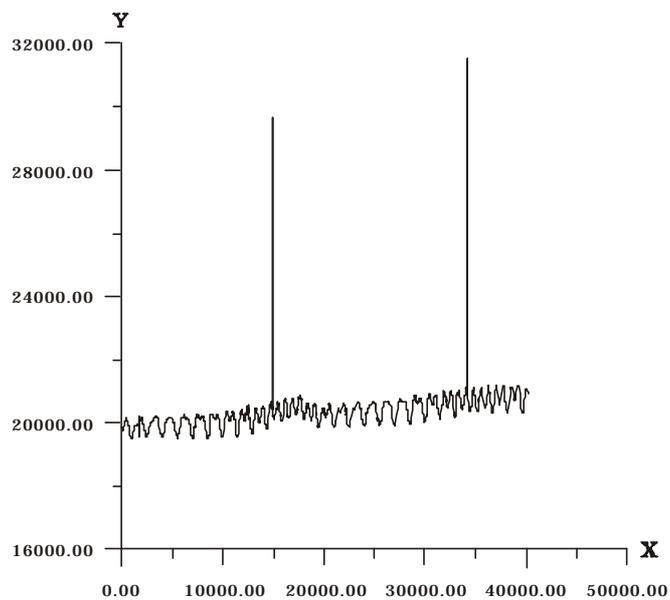

Fig.2

Readings of the gravimeter from Feb. 24,1996, to March 22,1996, inclusively.
$y$ is displacement of platinum weight. One division is *0.1 mm*, or *0.2 mgal*, $x$ is time in minutes.

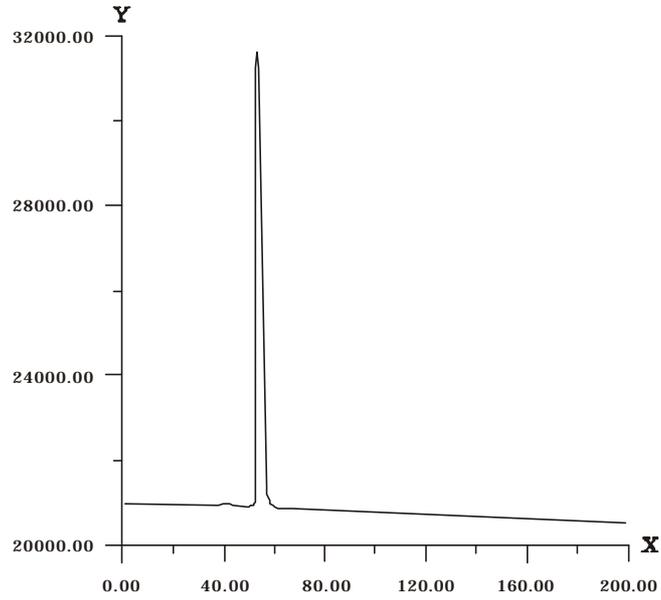

Fig.3

The event on March 18,1996.
*y* is displacement of platinum weight. One division is *0.1 mm*, or *0.2 mgal*, *x* is time in minutes.

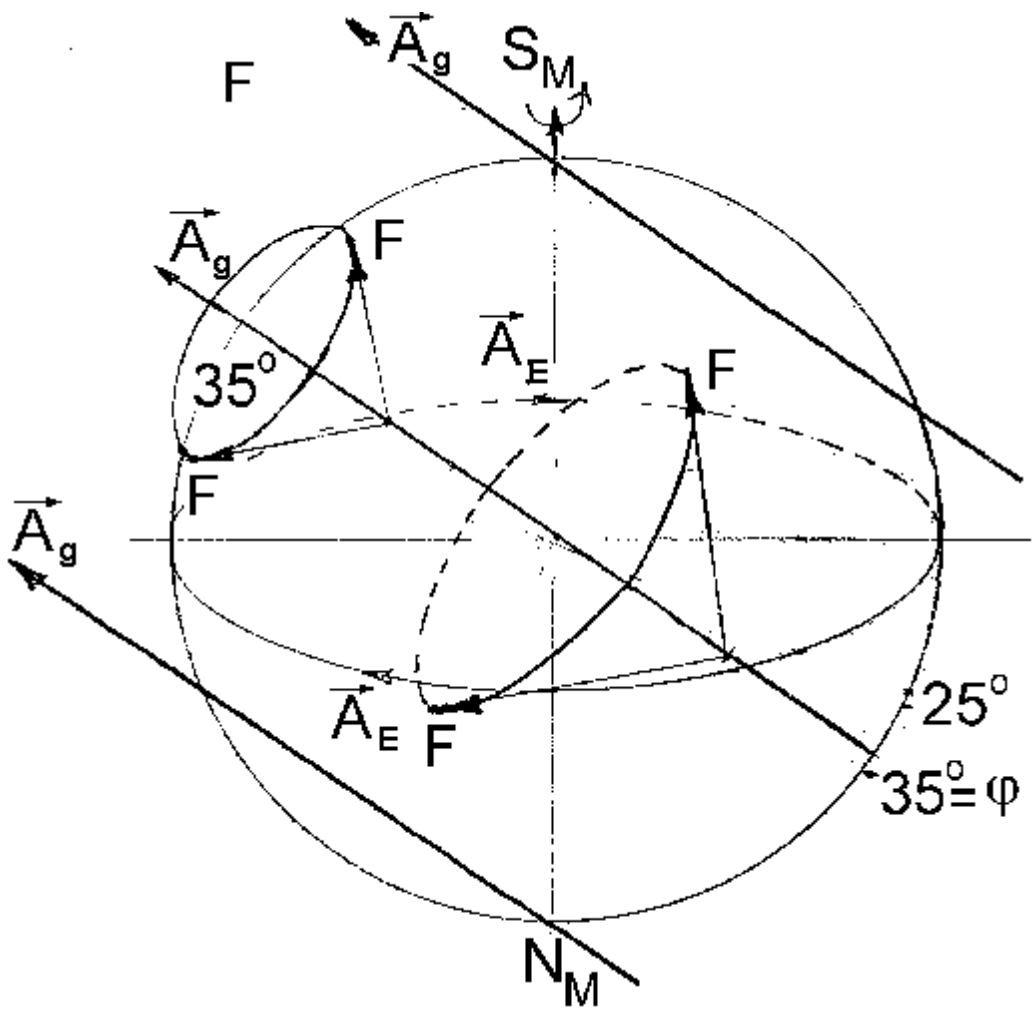

Fig.4

The direction of vector $\vec{A}_g$ and the cone of new force action (**F**) relatively to the globe.
$A_E$ – vectorial potential of Earth's magnetic field;
$S_M$, $N_M$ – South and North Pole of Earth's magnetic field.

Table2. The dependence of the distribution of arcs $Z_{Ag}$ on the latitude ö of earthquakes epicenter

| φ° | $Z_{ÃÄ}$ / n | 10° | | 30° | | 50° | | 70° | | 90° | | 110° | | 130° | | 150° | | 170° | | $\bar{n}$ | L | |
|---|---|---|---|---|---|---|---|---|---|---|---|---|---|---|---|---|---|---|---|---|---|---|
| | | | | | | | | | | | | | | | | | | | | | 0,1% | 1% |
| 55° | 128 | 0 0 | 0,00 0,00 | 22 22 | 1,55 1,55 | 26 26 | 1,83 1,83 | 21 31 | 1,48 2,18 | 57 46 | 4,01 3,24 | 2 3 | 0,14 0,21 | 0 0 | 0,00 0,00 | 0 0 | 0,00 0,00 | 0 0 | 0,00 0,00 | 14,2 | 3,1 2,9 | 2,4 2,2 |
| 45° | 155 | 9 4 | 0,52 0,23 | 24 26 | 1,40 1,51 | 28 27 | 1,63 1,57 | 24 30 | 1,40 1,74 | 34 38 | 1,98 2,21 | 36 30 | 2,09 1,74 | 0 0 | 0,00 0,00 | 0 0 | 0,00 0,00 | 0 0 | 0,00 0,00 | 17,2 | 2,4 2,4 | 1,9 2,0 |
| 35° | 159 | 19 19 | 1,07 1,07 | 21 25 | 1,19 1,44 | 21 20 | 1,19 1,13 | 24 23 | 1,36 1,30 | 33 38 | 1,86 2,75 | 41 34 | 2,32 1,92 | 0 0 | 0,00 0,00 | 0 0 | 0,00 0,00 | 0 0 | 0,00 0,00 | 17,7 | 2,3 2,3 | 1,9 1,9 |
| 25° | 91 | 13 18 | 1,29 1,78 | 14 12 | 1,39 1,19 | 16 10 | 1,59 0,99 | 9 9 | 0,89 0,89 | 12 15 | 1,09 1,49 | 20 13 | 1,98 1,29 | 7 14 | 0,69 0,13 | 0 0 | 0,00 0,00 | 0 0 | 0,00 0,00 | 10,1 | 2,1 2,0 | 1,7 1,6 |
| 15° | 122 | 13 5 | 0,96 0,37 | 8 23 | 0,59 1,69 | 22 14 | 1,62 1,03 | 20 21 | 1,47 1,54 | 25 17 | 1,84 1,54 | 12 19 | 0,88 1,40 | 22 23 | 1,62 1,69 | 0 0 | 0,00 0,00 | 0 0 | 0,00 0,00 | 13,6 | 2,1 2,1 | 1,7 1,8 |
| 5° | 111 | 0 0 | 0,00 0,00 | 24 20 | 1,95 1,63 | 17 15 | 1,38 1,22 | 12 16 | 0,98 1,30 | 11 12 | 0,89 0,98 | 15 15 | 1,22 1,22 | 24 21 | 1,95 1,71 | 8 12 | 0,65 0,98 | 0 0 | 0,00 0,00 | 12,3 | 2,1 2,0 | 1,8 1,7 |
| -5° | 194 | 0 0 | 0,00 0,00 | 20 24 | 0,33 1,11 | 42 34 | 1,94 1,57 | 20 23 | 0,93 1,06 | 18 23 | 0,83 1,06 | 33 27 | 1,53 1,25 | 35 24 | 1,62 1,11 | 26 39 | 1,20 1,81 | 0 0 | 0,00 0,00 | 21,6 | 2,1 2,0 | 1,7 1,6 |
| -10° | 152 | 1 0 | 0,06 0,00 | 1 2 | 0,06 0,13 | 35 36 | 2,07 2,13 | 19 17 | 1,12 1,01 | 21 16 | 1,24 0,95 | 22 26 | 1,30 1,54 | 26 23 | 1,54 1,36 | 22 25 | 1,30 1,48 | 5 7 | 0,30 0,41 | 16,9 | 2,1 2,2 | 1,7 1,8 |
| -25° | 141 | 0 0 | 0,00 0,00 | 1 2 | 0,06 0,13 | 23 25 | 1,46 1,59 | 29 29 | 1,85 1,85 | 21 18 | 1,34 1,15 | 16 17 | 1,02 1,08 | 22 19 | 1,40 1,21 | 17 22 | 1,08 1,40 | 12 9 | 0,76 0,57 | 15,7 | 1,60 1,60 | 1,3 1,4 |
| -35° | 53 | 0 0 | 0,00 0,00 | 1 0 | 0,17 0,00 | 0 2 | 0,00 0,34 | 17 14 | 2,89 2,37 | 9 9 | 1,53 1,53 | 9 9 | 1,53 1,53 | 4 9 | 0,68 1,53 | 6 6 | 1,02 1,02 | 7 4 | 1,19 0,68 | 5,9 | 2,5 2,3 | 2,0 1,8 |
| -45° | 26 | 0 0 | 0,00 0,00 | 0 0 | 0,00 0,00 | 1 1 | 0,34 0,34 | 5 6 | 1,72 2,07 | 4 4 | 1,38 1,38 | 3 1 | 1,03 0,34 | 3 3 | 1,03 1,03 | 7 7 | 2,41 2,41 | 3 4 | 1,03 1,38 | 2,9 | 2,3 2,4 | 1,8 1,9 |
| -55° | 39 | 0 0 | 0,00 0,00 | 0 0 | 0,00 0,00 | 0 0 | 0,00 0,00 | 1 2 | 0,23 0,47 | 12 13 | 2,79 3,02 | 10 9 | 2,33 2,09 | 7 4 | 1,63 0,93 | 9 11 | 2,09 2,56 | 0 0 | 0,00 0,00 | 4,3 | 2,9 2,9 | 2,2 2,3 |
| $c_p$ α=270° | 114,25 | 4,58 | | 6,67 | | 19,25 | | 16,75 | | 21,42 | | 18,25 | | 12,50 | | 7,92 | | 2,42 | | | | |
| $c_p$ α=290° | 114,25 | 3,83 | | 13,00 | | 17,50 | | 18,42 | | 20,75 | | 16,92 | | 11,17 | | 10,17 | | 2,00 | | | | |
| $\Sigma_{270°}$ | 1371 | 55 | 0,36 | 80 | 0,53 | 231 | 1,52 | 201 | 1,32 | 257 | 169 | 219 | 1,44 | 150 | 0,99 | 95 | 0,62 | 29 | 0,19 | 152,3 | | |
| $\Sigma_{290°}$ | 1371 | 46 | 0,30 | 156 | 1,02 | 210 | 1,38 | 221 | 1,45 | 249 | 1,63 | 203 | 1,33 | 140 | 0,92 | 122 | 0,80 | 24 | 0,16 | 152,3 | | |

Table.3
Greatest earthquakes in the world in the second part of
the XX century and in XXI century.

| ¹ | Place, coordinates | Date, UTC time (Greenwich time) | Magnitude by Richter scale | á | â |
|---|---|---|---|---|---|
| 1 | Kamchatka **52.76N 160.06E** | 1952 November 04 16:58:26.0 | 9.0 | 1° | - |
| 2 | Andreanof Islands, Alaska **51.56N 175.39W** | 1957 March 09 | 8.6 | 90° | - |
| 3 | Chile **38.24S 73.05W** | 1960 May 22 19:11:14 | 9.5 | 12° | - |
| 4 | Prince William Sound, Alaska **61.02N 147.65E** | 1964 March 28 03:36:14 | 9.2 | 14° | - |
| 5 | Armenia **41.0 N 44.2 E** | 1988, December 7, 07:41 | 6,9 | - | 41° |
| 6 | West Coast of Northen Sumatra **3.295 N 95.982 E** | 2004, December 26 00:58 | 9.0 | 90° | - |
| 7 | Sakhalin Island **52.63 N 142.83 E** | 1995, May 27 13:03 | 6,7 | - | 40° |
| 8 | Pakistan **34.53 N 73.58E** | 2005 October 08 19:46 | 7.6 | 90° | - |

á - Angle between projection of the vector Zenith and projection of the vector $A_g$ on the plane of the ecliptic.
â - Angle between projection of the vector V and projection of the vector $A_g$ on the plane of the ecliptic.

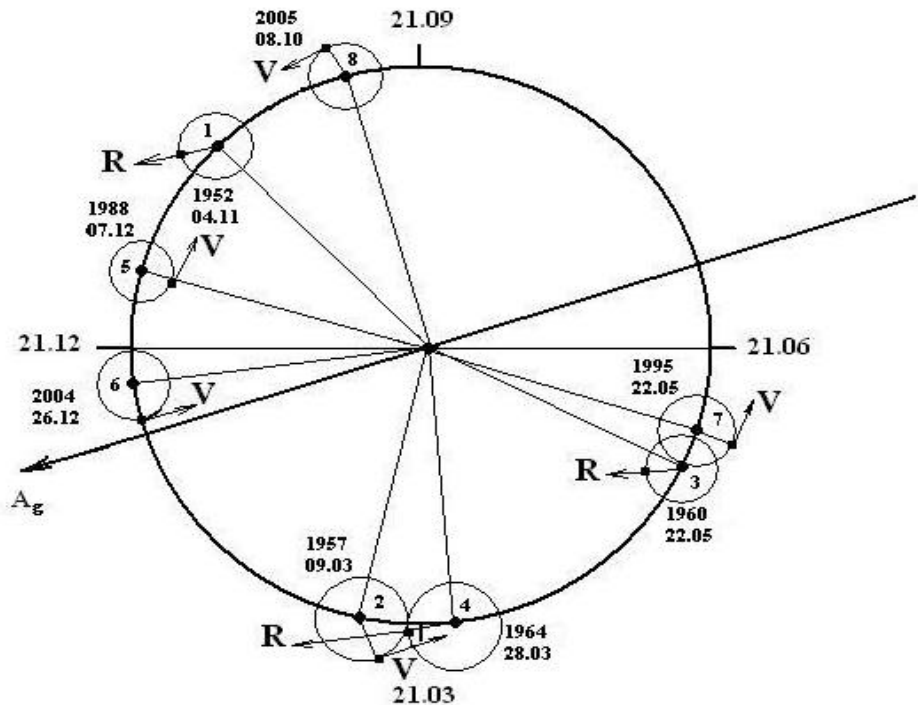

Fig.5
Directions of vectors: $A_g$, V and R (radius vector from Earth center or vector Zenith) for **g**reatest earthquakes in the world in the second part of
the XX century and in XXI century (Table. 3).